\documentclass[preprint,nofootinbib,showpacs,prd,aps]{revtex4-1}
\usepackage{amsmath,amsfonts,amssymb,latexsym}
\usepackage{amsmath}
\usepackage{amsfonts}
\usepackage{amssymb}
\usepackage{graphicx}

\def \be{\begin{equation}}
\def \ee{\end{equation}}
\def \bq{\begin{eqnarray}}
\def \eq{\end{eqnarray}}
\def \beq{\begin{eqnarray*}}
\def \eeq{\end{eqnarray*}}

\begin{document}

\title{{\Large The Generic  Sudden Singularity in Brans-Dicke Theory}}
\author{John D. Barrow\footnote{Deceased}}
\affiliation{DAMTP, Centre for Mathematical Sciences \\
Wilberforce Rd., Cambridge University \\
Cambridge CB3 0WA, UK}
\author{Spiros Cotsakis}
\email{skot@aegean.gr}
\affiliation{Institute of Gravitation and Cosmology\\ RUDN University\\
ul. Miklukho-Maklaya 6, Moscow 117198\\ Russia}
\affiliation{Research Laboratory of Geometry, \\ Dynamical Systems and Cosmology\\
University of the Aegean\\ Karlovassi 83200, Samos\\ Greece}
\author{Dimitrios Trachilis}
\email{dimitrios.trachilis@aum.edu.kw}
\affiliation{Department of Mathematics\\
College of Engineering and Technology \\
American University of the Middle East\\
Kuwait}
\begin{abstract}
\noindent We construct a formal asymptotic series expansion for a general solution of the Brans-Dicke equations with a fluid source near a sudden singularity. This solution contains eleven independent arbitrary functions of the spatial coordinates as required by the Cauchy problem of the theory. We  show that the solution is geodesically complete and has the character of a shock wave in the sudden  asymptotic region. This solution is weak in the senses of Tipler and Krolak as in the corresponding case of general relativity.
\end{abstract}
\date{December, 2020}
\maketitle
\newpage
\tableofcontents
\newpage
\section{Introduction}
Formal series expansions in powers of the proper time has been a central technique associated with existence theorems for the Einstein equations and the Cauchy problem since the very beginning of mathematical relativity \cite{ycb09}. For most of these studies, the relevant expansions where built around regular, as opposed to singular, spacetime points, and this was particularly true for the local Cauchy problem. For global problems of solutions of the Einstein equations one is faced with the issue of singularities and the related problems of the domain of validity of the solutions so constructed \cite{i}, \cite{c}.

Directly related to the study of formal expansions is the issue of \emph{function-counting} which provides a count of the independent equations and unknowns needed to have a determinate problem.  Most importantly, function counting is needed in order to have a characterization of whether or not the studied solutions have the required number of arbitrary functions to qualify as generic, that is to be general solutions of the problem. So far as one restricts to formal expansions which solve the field equations (in the sense of direct substitution of these infinite series in the equations leading to all terms cancelling), without further demanding that the series is convergent or even asymptotic, does not strictly require that such expansions be constructed around regular points in spacetime. This does not mean that formal expansions around singular points are necessarily divergent ones. In fact, one does not even have to restrict to integer exponents in the expansions, but may allow for fractional ones or even consider more general Fuchsian formal series \cite{k}, \cite{gor}. For example, the original quasi-isotropic constructions of Refs. \cite{LL}-\cite{KL2} have the property that they describe asymptotic forms of special solutions that do not contain the required number of arbitrary functions to qualify as general solutions in vacuum, or containing one or more fluids \cite{starob}, \cite{starob2}.

This interplay between general and special solutions in gravitation and cosmology is not new, and in fact this `generality problem' is intimately connected not only with the function-counting issue but also with other basic problems such as the singularity, isotropization and cosmic no-hair problems (cf. \cite{ba17} and references therein). There are four known applications of function-counting that lead to the construction of generic solutions, and identify those that have the necessary maximum number of free functions and are general solutions of the equations. The first corresponds to the problem of the Einstein equations with positive cosmological constant or with a $p=-\rho$ fluid, and builds solutions around the de Sitter solution  which comply with the expectations of the no-hair theory \cite{no hair}-\cite{rend}. The second result concerns the behaviour of the Einstein equations with an `ultrastiff perfect fluid' having $p>\rho$ near quasi-isotropic singularities by Heinzle and Sandin \cite{hs}. The third result is a  generic solution that describes the approach to a `sudden' finite-time singularity  of Barrow, Cotsakis and Tsokaros in Ref. \cite{ bct}, which has the property that the approaching solution is geodesically complete  and the discontinuities produce a weakly singular solution cf. \cite{bc13}. The fourth known result which leads to a generic solution in terms of function counting is a vacuum analytic  solution of $R+\epsilon R^2$ gravity constructed in \cite{ct16} which represents an asymptotic formal series expansion.

In this paper we develop and study a further application of the function-counting method to Brans-Dicke theory, and construct the first  generic solution near a sudden singularity in this context. By examining the behaviour of geodesics, we show that the generic approach to the sudden region may be given as a shock wave to both general relativity and the Brans-Dicke theory, and also that solutions represent weak singularities. These results add to the physical relevance of sudden singularities.   The latter have been widely studied since their introduction in Refs. \cite{jdb04aa}-\cite{jdb04b} as asymptotic  regions in spacetime were matter satisfies the strong energy condition, the scale factor and its first derivative are finite, but discontinuities occur in its second derivative and fluid pressure. The stability of these solutions to small perturbations has been shown using a gauge invariant formalism in Ref. \cite{lip}, and have also been shown to be stable against quantum particle production in \cite{fab}. Their description using fractional formal series has been given in Refs. \cite{scot}, \cite{bct1}. In the context of Brans-Dicke theory which generalizes general relativity by allowing for the possibility of the Newtonian gravitational constant $G$, sudden singularities were first discovered in Ref. \cite{b19} where it was shown that they possess a number of interesting properties.

The plan of this paper is as follows. In the next Section we give a review of the most important properties of the simplest solution with a sudden singularity discovered in \cite{b19}. From the property of simultaneity of this solution with a sudden singularity  we estimate the end values of the scale factor and the  Brans-Dicke scalar field for large values of the coupling parameter $\omega$.  In Section 3, we give the formal expansions near the sudden singularity for the spatial metric and its inverse and compare with similar expansions for non-sudden regions. We also provide a discussion of the homogeneous vs. inhomogeneous scalar field expansions in the present context, which  as they turns out are closely related to the singularity being simultaneous or not. A detailed derivation of the theory for the case of a homogeneous scalar field expansion is given in Appendix A. In Section 4, we provide the expansions for the various curvatures, while in Section 5 we find the splittings and develop the process of balancing of the various terms for the field equations. These provide the necessary information for the various unknowns of the problem and their functional relations, so that a final counting of the arbitrary functions present in the solution can be given in Section 6. Section 7 gives a further application of these results to the geodesic behaviour at the sudden singularity, and the ensuing interpretation of it as a shock wave. We also  provide conditions for the sudden singularity to be a weak one in Brans-Dicke theory. We discuss our findings in Section 8. Our notation is that of Ref. \cite{LL}.

\section{The simplest sudden singularity in Brans-Dicke}
In this Section, we give a brief review of the most important properties of the first and simplest  solution with a sudden singularity discovered in Brans-Dicke theory in Ref. \cite{b19}. The Brans-Dicke equations in the standard form read,
\begin{eqnarray}
R_{ij} - \frac{1}{2}g_{ij}R &=& \frac{8\pi}{\phi}T_{ij} + \frac{\omega}{\phi^2}(\partial_i \phi \partial_j \phi -
\frac{1}{2}g_{ij}\partial^k \phi \partial_k \phi) + \frac{1}{\phi}(\nabla_i \nabla_j \phi - g_{ij}\Box_g \phi),
\label{G_ij} \label{bd}\\
\Box_g \phi &=& \frac{8\pi}{3+2\omega}T,
\label{Boxphi} \\
\nabla_i T^i_j &=& 0.
\label{conlaws}
\end{eqnarray}
\subsection{Form of the solution}
Following \cite{b19}, we assume the spacetime metric to be homogeneous and isotropic with an FRW line element $ds^2=dt^2-a^2 (t)d\Omega^2$, with $d\Omega^2$ denoting the metric of spatial sections. Then ordering the terms in increasing powers of the proper time $t$,  the solution with a sudden singularity in this context is given by,
\begin{eqnarray}\label{bdss1}
a(t) &=& \left(\frac{t}{t_{s}}\right)^{q}\left( a_{s}-1\right) +1-\left(1-\frac{t}{t_{s}}\right)^{n} \label{sol2}, \\
\phi(t) &=& \phi_s \left(\frac{t}{t_s}\right)^r - \frac{3}{\omega a_s}\left(1 - \frac{t}{t_s}\right)^n. \label{solphi}\label{bdss2}
\end{eqnarray}%
This solution requires $0<r<1<n<2$ and $0<q\leq 1$,  $a_{s}\equiv a(t_{s})$ and $\phi_{s}\equiv \phi(t_{s})$ are the (finite) values of the scale factor and scalar field at the sudden singularity, and the coefficient in the second term in Eq. (\ref{bdss2}) is required for consistency in order to assure simultaneity at the sudden singularity. As shown in \cite{b19}, we can work with the \emph{flat} FRW metric without loss of generality because the curvature terms are finite at the sudden singularity, hence subdominant with respect to other terms which diverge there.

The solution exists on the interval $0<t<t_{s}$.  `Early' in this context means `as $t$ approaches zero from above', while the sudden singularity time $t_s$ appears as a future singularity. Another way to view the evolution of a universe with a future sudden singularity is to measure the time as a fraction of the `sudden singularity duration' $t_s$ up to the sudden singularity,  and after it as a multiple of $t_s$. Since the sudden singularity occurs at $t_s$ which is an arbitrary positive constant,  we can divide by it and measure the total time $t$ in terms  of $t_s$. In this sense, the  interval of existence of the solution up to the sudden singularity is $(0,1)$, and the solution has a simpler appearance: $a=(a_{s}-1)t^q+1-(1-t)^n,\,\, \phi=\phi_s t^r-\frac{3}{\omega a_s}(1-t)^n$.
\subsection{Properties of the FRW sudden singularity}
As shown in Ref. \cite{b19}, the sudden singularity  solution (\ref{bdss1}), (\ref{bdss2}) possesses a number of properties, and we focus below on the most important ones for our current purposes.
\begin{itemize}
\item The sudden singularity  is necessarily simultaneous, occuring at proper time $t_s$, for all diverging fields, namely,  $\ddot{a},\ddot{\phi}, \dot{\rho},p$.
\item The generality of non-Machian  (scalar field dominates over matter) over `special' Machian solutions (matter dominates over $\phi$) as $t\downarrow 0$, leads us to prefer initial conditions of the form $\phi(0)\neq 0$.
\item The previous two properties continue to hold for more general scalar-tensor theories when certain restrictions on the Brans-Dicke  function $\omega(\phi)$ and scalar field potential $V(\phi)$ are imposed.
\end{itemize}
\subsection{Estimate of the scale factor}
In the remaining of this Section, we provide an estimate of the boundary values of the fields at the sudden singularity of the Brans-Dicke theory for large values of the coupling parameter $\omega$. This estimate is a consequence of the property of simultaneity of the sudden singularity discussed above.

The initial and final limiting values of the sudden singularity solution read,
\begin{eqnarray}
\lim_{t\rightarrow 0}a(t) &=& 0,\quad \lim_{t\rightarrow 1}a(t)=a_s \label{in1}, \\
\lim_{t\rightarrow 0}\phi(t) &=& - \frac{3}{\omega a_s},\quad \lim_{t\rightarrow 1}\phi(t) = \phi_s. \label{in2}
\end{eqnarray}
In the case of non-Machian solutions, and assuming that the sudden singularity is simultaneous, then since the constant $a_s$ is arbitrary, we can set,
\be
a_s=\frac{3}{\omega},
\ee
so that $\phi(0)=-1$, as required. In this case, the size of the scale factor at the sudden singularity, $a_s$, is related inversely proportional to the Brans-Dicke coupling constant $\omega$.

As $\omega$ becomes large and the limit to the theory of general relativity is approached, the scale factor size becomes very small at the sudden singularity. This is in accordance with the asymptotic behaviour found in Ref. \cite{b19}, Eq. (29).

This result may be interpreted as an upper bound estimate, coming the Brans-Dicke theory, of the size of the scale factor appearing in the \emph{general relativistic} sudden singularity solution  found in Ref. \cite{jdb04a}.

Using the isotropic and homogeneous solution (\ref{bdss1}), (\ref{bdss2}) as a guide, in the next Section we
build a series expansion of the metric in the neighbourhood of the sudden
singularity, as a first step of the program, set out in this work, to construct the most general cosmological solution to the Brans-Dicke equations having a sudden singularity. It will be of interest to examine which of the properties of the simplest solution of the Brans-Dicke equations with a sudden singularity considered in the present Section, pass over to that more general situation.

\section{Metric series expansions}
\subsection{Formal expansions for the metric and its inverse}
The general form of the metric in
synchronous coordinates is (Latin indices are for spacetime
components, Greek for spatial ones),
\begin{equation}
ds^{2}=dt^{2}-\gamma_{\alpha \beta}dx^{\alpha}dx^{\beta}.
\label{met}
\end{equation}%
The future sudden singularity is approached as $t\rightarrow 1$, and the solutions (\ref{bdss1}), (\ref{bdss2}) have the  asymptotic forms (\ref{in1}), (\ref{in2}),
\be
a(t) \rightarrow a_{s},\quad  \phi(t) \rightarrow \phi_s. \label{lin}
\ee
So the expansion of the spatial metric in Eq.  (\ref{met}) and Brans-Dicke scalar field $\phi$ near the sudden singularity must start with constant terms, and have the next few terms in the form,
\bq
\gamma _{\alpha \beta }&=&a_{_{\alpha \beta }}+b_{_{\alpha \beta
}}t+c_{_{\alpha \beta }}t^{n}+ O(t^2),\label{seriesgab}\\
\phi &=& \phi_0 + \phi_1 t + \phi_n t^{n} + O(t^2)\label{seriesphi}.
\eq
Here $1<n<2$, and the coefficients in the formal series are inhomogeneous functions of the space coordinates,
\be
\phi_i=\phi_i(x), \quad a_{\alpha \beta }=a_{\alpha \beta }(x),b_{\alpha \beta }=b_{\alpha \beta }(x), c_{\alpha \beta}=c_{\alpha \beta}(x).
\ee
The  final form of the sudden singularity expansion of the \emph{ inverse} metric tensor reads:
\begin{equation}
\gamma ^{\alpha \beta }=a^{\alpha \beta }-b^{\alpha \beta }\;t-c^{\alpha
\beta }\;t^{n} + O(t^2),
\label{invgab}
\end{equation}%
where $\gamma _{\alpha \beta }\gamma ^{\beta \gamma }=\delta _{\alpha
}^{\gamma }$, and  $a_{\alpha \beta }a^{\beta \gamma }=\delta
_{\alpha }^{\gamma }$. Also here the indices of $b_{\alpha \beta }$ and $c_{\alpha \beta}$ are raised by $a^{\alpha \beta }$.

It is very instructive to compare the difficulty of the calculations introduced by a sudden singularity to the non-sudden ones, and so we shall include below some of the many proofs of key results. To prove Eq. (\ref{invgab}), we start by the spatial metric expansion,
\begin{equation}
\gamma _{\alpha \beta }=a_{_{\alpha \beta }}+b_{_{\alpha \beta
}}t+c_{_{\alpha \beta }}t^{n}+ d_{_{\alpha \beta }}t^{2} +\cdots,
\end{equation}
where we include the term of order two so as to give explicitly the corresponding term for the inverse metric expansion. The problem here is to start with a general expansion of the form,
\begin{equation}
\gamma ^{\alpha \beta }=\widetilde{a}^{\alpha \beta }t^p + \widetilde{b}^{\alpha \beta }t^q
+\widetilde{c}^{\alpha \beta }t^r + \widetilde{d}^{\alpha \beta }t^s +\cdots, \qquad p<q<r<s,
\end{equation}
 and calculate the coefficients and exponents for each term.
By definition we have,
\begin{equation}
\gamma ^{\alpha \mu }\gamma _{\mu \beta }=\delta ^{\alpha}_{\beta},
\end{equation}
and so substituting the expansions,
\begin{equation}
(\widetilde{a}^{\alpha \mu }t^p + \widetilde{b}^{\alpha \mu }t^q
+\widetilde{c}^{\alpha \mu }t^r + \widetilde{d}^{\alpha \mu }t^s +\cdots) \times
(a_{_{\mu \beta }}+b_{_{\mu \beta}}t+c_{_{\mu \beta }}t^{n}+ d_{_{\mu \beta }}t^{2} +\cdots)=\delta ^{\alpha}_{\beta},
\end{equation}
we arrive at the \emph{balancing conditions},
\begin{eqnarray}
&& \widetilde{a}^{\alpha \mu}a_{\mu \beta}t^p + \widetilde{a}^{\alpha \mu}b_{\mu \beta}t^{p+1}
+ \widetilde{a}^{\alpha \mu}c_{\mu \beta}t^{p+n} + \widetilde{a}^{\alpha \mu}d_{\mu \beta}t^{p+2} + \nonumber \\
&& \widetilde{b}^{\alpha \mu}a_{\mu \beta}t^q + \widetilde{b}^{\alpha \mu}b_{\mu \beta}t^{q+1}
+ \widetilde{b}^{\alpha \mu}c_{\mu \beta}t^{q+n} + \widetilde{b}^{\alpha \mu}d_{\mu \beta}t^{q+2} + \nonumber \\
&& \widetilde{c}^{\alpha \mu}a_{\mu \beta}t^r + \widetilde{c}^{\alpha \mu}b_{\mu \beta}t^{r+1}
+ \widetilde{c}^{\alpha \mu}c_{\mu \beta}t^{r+n} + \widetilde{c}^{\alpha \mu}d_{\mu \beta}t^{r+2} + \nonumber \\
&& \widetilde{d}^{\alpha \mu}a_{\mu \beta}t^s + \widetilde{d}^{\alpha \mu}b_{\mu \beta}t^{s+1}
+ \widetilde{d}^{\alpha \mu}c_{\mu \beta}t^{s+n} + \widetilde{d}^{\alpha \mu}d_{\mu \beta}t^{s+2} + \cdots
= \delta ^{\alpha}_{\beta}.
\end{eqnarray}
These are the necessary equations from which to infer the forms of the various unknowns.
The lowest exponent in the series is $p$, so balancing this term first we find  $p=0$, and,
\begin{equation}
\widetilde{a}^{\alpha \mu }a_{\mu \beta}=\delta ^{\alpha}_{\beta},
\end{equation}
which means that the coefficient is simply,
\begin{equation}
\widetilde{a}^{\alpha \beta }=a^{\alpha \beta}.
\end{equation}
Next lowest order is $t^{p+1}=t$ or $t^q$. If $q\neq 1$, then
$\widetilde{a}^{\alpha \mu }b_{\mu \beta}=a^{\alpha \mu }b_{\mu \beta}=b^\alpha _\beta =0$. So $q=1$, and hence,
\begin{equation}
0 = \widetilde{a}^{\alpha \mu }b_{\mu \beta} + \widetilde{b}^{\alpha \mu }a_{\mu \beta} =
a^{\alpha \mu }b_{\mu \beta} + \widetilde{b}^{\alpha \mu }a_{\mu \beta} =
b^\alpha _\beta + \widetilde{b}^{\alpha \mu }a_{\mu \beta}.
\end{equation}
Then,
\begin{equation}
0 = b^\alpha _\beta a^{\beta \epsilon} + \widetilde{b}^{\alpha \mu }a_{\mu \beta}a^{\beta \epsilon} =
b^{\alpha \epsilon} + \widetilde{b}^{\alpha \mu }\delta ^\epsilon _\mu =
b^{\alpha \epsilon} + \widetilde{b}^{\alpha \epsilon},
\end{equation}
namely that,
\begin{equation}
\widetilde{b}^{\alpha \beta} = -b^{\alpha \beta}.
\end{equation}
Next lowest order is $t^n$ or $t^r$. If $r\neq n$, then
$\widetilde{a}^{\alpha \mu }c_{\mu \beta}=a^{\alpha \mu }c_{\mu \beta}=c^\alpha _\beta =0$. So $r=n$, thus,
\begin{equation}
0 = \widetilde{a}^{\alpha \mu }c_{\mu \beta} + \widetilde{c}^{\alpha \mu }a_{\mu \beta} =
a^{\alpha \mu }c_{\mu \beta} + \widetilde{c}^{\alpha \mu }a_{\mu \beta} =
c^\alpha _\beta + \widetilde{c}^{\alpha \mu }a_{\mu \beta}.
\end{equation}
Then,
\begin{equation}
0 = c^\alpha _\beta a^{\beta \epsilon} + \widetilde{c}^{\alpha \mu }a_{\mu \beta}a^{\beta \epsilon} =
c^{\alpha \epsilon} + \widetilde{c}^{\alpha \mu }\delta ^\epsilon _\mu =
c^{\alpha \epsilon} + \widetilde{c}^{\alpha \epsilon},
\end{equation}
so that,
\begin{equation}
\widetilde{c}^{\alpha \beta} = -c^{\alpha \beta}.
\end{equation}
Proceeding in a similar fashion, calculations become somewhat more complicated for the higher-order terms.  The next lowest order is $t^s$ or $t^2$. If $s\neq 2$, then
$\widetilde{a}^{\alpha \mu }d_{\mu \beta}=a^{\alpha \mu }d_{\mu \beta}=d^\alpha _\beta =0$. So $s=2$, thus,
\begin{equation}
0 = \widetilde{a}^{\alpha \mu }d_{\mu \beta} + \widetilde{b}^{\alpha \mu }b_{\mu \beta} +
\widetilde{d}^{\alpha \mu }a_{\mu \beta} =
a^{\alpha \mu }d_{\mu \beta} - b^{\alpha \mu }b_{\mu \beta} +
\widetilde{d}^{\alpha \mu }a_{\mu \beta} =
d^\alpha _\beta - b^{\alpha \mu }b_{\mu \beta} + \widetilde{d}^{\alpha \mu }a_{\mu \beta}.
\end{equation}
Then,
\begin{equation}
\widetilde{d}^{\alpha \mu }a_{\mu \beta} = b^{\alpha \mu }b_{\mu \beta} - d^\alpha _\beta.
\end{equation}
So,
\begin{equation}
\widetilde{d}^{\alpha \mu }a_{\mu \beta}a^{\beta \epsilon} =
b^{\alpha \mu }b_{\mu \beta}a^{\beta \epsilon} - d^\alpha _\beta a_{\mu \beta}a^{\beta \epsilon},
\end{equation}
thus,
\begin{equation}
\widetilde{d}^{\alpha \mu }\delta ^\epsilon _\mu =
b^{\alpha \mu }b^\epsilon _\mu - d^{\alpha \epsilon}.
\end{equation}
Finally, we find that,
\begin{equation}
\widetilde{d}^{\alpha \beta } = b^{\alpha \gamma }b^\beta _\gamma - d^{\alpha \beta}.
\end{equation}
Therefore,
\begin{equation}
\gamma ^{\alpha \beta }=a^{\alpha \beta } - b^{\alpha \beta }t
- c^{\alpha \beta }t^n + (b^{\alpha \gamma }b^\beta _\gamma - d^{\alpha \beta })t^2 +\cdots,
\end{equation}
and the proof  of the inverse metric expansion is now complete.

The order of the field equations introduced later will be equal to two, and we further need only be concerned with spatial metric expansions up to order $n$ for the present problem. Including terms of second and higher orders in the metric and scalar field formal expansions is equivalent to extra constant or vanishing terms appearing in the various curvatures  at the sudden singularity in addition to other terms which involve $n$ and are already diverging. Such terms are subdominant near the sudden singularity with respect to terms of order involving $n$. This is a new feature that is only present in sudden singularity expansions, absent in the more conventional problem of Taylor-like expansions in non-sudden types.

\subsection{Homogeneous vs. inhomogeneous expansions}
It is an interesting question whether we can proceed without loss of generality by considering a simpler type of formal expansion for the metric and the scalar field. This could be like the above expansions (\ref{seriesgab}), (\ref{seriesphi}) but \emph{the scalar field coefficients  be only constants, not purely spatial functions}. Although there is no reason for restricting the expansions to such `homogeneous' ones, this special case is treated fully in the Appendix. As it turns out,  the final result is of the same generality as that obtained by the general inhomogeneous expansions treated in the main body of this paper.

However, there is an important difference in the interpretation of the two cases: when non-Machian solutions are sought for (non-zero constant first term in the formal series for the metric), the simpler case containing constant coefficients for the scalar field leads to the generic sudden singularity being \emph{necessarily} a simultaneous one, in distinction to the general \emph{in}homogeneous case where the diverging terms are generically non-simultaneous at the sudden singularity.

We believe this effect has to do with the fact that in this problem the inhomogeneous solution having a sudden singularity in the context of the Brans-Dicke equations (\ref{G_ij}), (\ref{Boxphi}), (\ref{conlaws}) turns out to be a general one in the sense of function counting, as we prove below.

\section{Curvature expansions}
In this Section, we build the necessary formal expansions for the connection and for the various curvatures near the sudden singularity. These synchronous system calculations  are  sometimes very lengthy due to the form of the unperturbed solution, but also (later) due to the inclusion of the Brans-Dicke scalar field.

\subsection{Extrinsic curvature}
We assume a non-null hypersurface $\Sigma$ with normal vector $n_\alpha$. The covariant extrinsic curvature expansions as we approach the sudden singularity  are straightforward to give,
\be
K_{\alpha \beta }=\partial_t \gamma _{\alpha \beta }= b_{\alpha \beta }+ nc_{\alpha \beta }\;t^{n-1}+ 2d_{\alpha \beta}t + \cdots.
\ee
Then the mixed tensor is calculated as follows,
\bq
K^{\beta }_{\alpha } &=& \gamma^ {\beta \mu} K_{\mu \alpha} \nonumber \\
&=& [a^{\beta \mu } - b^{\beta \mu }t - c^{\beta \mu }t^n + (b^{\beta \gamma }b^\mu _\gamma - d^{\beta \mu })t^2 +\cdots] \times  [b_{\mu \alpha }+ nc_{\mu \alpha}t^{n-1}+ 2d_{\mu \alpha }t + \cdots] \nonumber \\
&=& b^{\beta }_{\alpha }+nc^{\beta }_{\alpha}\;t^{n-1}+ (2d^\beta_\alpha - b^\beta _\gamma b^\gamma _\alpha)t
+(-n b^\beta_\gamma c^\gamma_\alpha - b^\gamma_\alpha c^\beta_\gamma)t^n\nonumber\\
 &+& (-2c^\beta_\gamma d^\gamma_\alpha - n c^\gamma_\alpha d^\beta_\gamma +
n b^{\beta \gamma}b^\mu_\gamma c_{\mu \alpha})t^{n+1} + \cdots.\label{kab}
\eq
Tracing, we find,
\bq\label{k}
K = K^\alpha_\alpha &=& b+nc\;t^{n-1} + (2d-b^\beta_\alpha b^\alpha_\beta)t - (n+1)b^\beta_\alpha c^\alpha_\beta t^n\nonumber\\
&+&\left(-(n+2)c^\beta_\alpha d^\alpha_\beta + n b^\beta_\alpha b^\alpha_\gamma c^\gamma_\beta\right)t^{n+1}+ \cdots .
\eq
The trace $K=\nabla_\alpha n^\alpha$ of the extrinsic curvature has of course the standard interpretation as the expansion of a congruence of (non-null) geodesics that intersect the hypersurface $\Sigma$ orthogonally (with tangent vector $n_\alpha$ at $\Sigma$). However, these expansions for the extrinsic curvature are in addition of an asymptotic nature in the present case, and that any statement involving $K$ becomes automatically an asymptotic one. For instance, a diverging (resp. converging) congruence of geodesics orthogonal to $\Sigma$ having $K>0$ (resp. $K<0$) are examples of such asymptotic statements. This asymptotic interpretation will play an important role below.

From (\ref{kab}), we find by taking the (proper) time derivative,
\bq
\partial_t K^{\beta }_{\alpha } &=& n(n-1)c^{\beta }_{\alpha}\;t^{n-2} + (2d^\beta_\alpha - b^\beta _\gamma b^\gamma _\alpha)
+n(-n b^\beta_\gamma c^\gamma_\alpha - b^\gamma_\alpha c^\beta_\gamma)t^{n-1} \nonumber \\
&+& (n+1)(-2c^\beta_\gamma d^\gamma_\alpha - n c^\gamma_\alpha d^\beta_\gamma +
n b^{\beta \gamma}b^\mu_\gamma c_{\mu \alpha})t^n + \cdots ,
\eq
whereas the time derivative for the evolution of the expansion $K$ from Eq. (\ref{k}) is given by,
\bq\label{dotk}
\partial_t K &=&n(n-1)c\;t^{n-2} + (2d-b^\beta_\alpha b^\alpha_\beta) - n(n+1)b^\beta_\alpha c^\alpha_\beta t^{n-1}
\nonumber \\
&+&(n+1)[-(n+2)c^\beta_\alpha d^\alpha_\beta + n b^\beta_\alpha b^\alpha_\gamma c^\gamma_\beta]t^n +\cdots .
\eq
An important conclusion follows directly from these results. As we approach the sudden singularity at zero proper time, we have from Eqns.
(\ref{k}), (\ref{dotk}),
\be
\lim_{t\rightarrow 0}K = b,\quad\lim_{t\rightarrow 0}\partial_t K=\textrm{sign}(c)\times \infty.
\ee
This will be used later.

Below we shall also make use of the products, found by direct use of the relations above:
\bq
K_{\alpha }^{\beta }K_{\beta }^{\alpha } &=& b_{\alpha }^{\beta }b^{\alpha }_{\beta }+2nb_{\alpha }^{\beta }c^{\alpha }_{\beta } \;t^{n-1} + (4b^\beta_\alpha d^\alpha_\beta - 2b^\beta_\alpha b^\alpha_\gamma b^\gamma_\beta)t + \cdots , \\
KK^{\beta }_{\alpha } &=& bb^{\beta }_{\alpha }+n(cb^{\beta }_{\alpha}+bc^{\beta }_{\alpha })\;t^{n-1} + (2bd^\beta_\alpha + 2db^\beta_\alpha  - bb^\beta_\gamma b^\gamma_\alpha - b^\beta_\alpha b^\gamma_\delta b^\delta_\gamma)t + \cdots.
\end{eqnarray}

\subsection{Ricci curvature}
In a synchronous reference system,  the 4-dimensional Christoffel symbols, namely,
\begin{equation}
\Gamma^i_{jk}=\frac{1}{2}g^{il}(\partial_k g_{lj} + \partial_j g_{lk} - \partial_l g_{jk}),
\label{Chrostffel}
\end{equation}
have  components, as defined below, which satisfy,
\begin{eqnarray}
\Gamma^0_{00} &=& \Gamma^\alpha_{00} = \Gamma^0_{0 \alpha} = 0, \\
\Gamma^0_{\alpha \beta} &=& \frac{1}{2}K_{\alpha \beta} = \frac{1}{2}b_{\alpha \beta} + \frac{1}{2}n c_{\alpha \beta} t^{n-1} + O(t), \\
\Gamma^\alpha_{0 \beta} &=& \frac{1}{2}K^\alpha_\beta = \frac{1}{2}b^\alpha_\beta + \frac{1}{2}n c^\alpha_\beta t^{n-1} + O(t), \\
\Gamma^\alpha_{\beta \gamma} &=& \lambda^\alpha_{\beta \gamma} = \frac{1}{2}\gamma^{\alpha \delta}(\partial_\gamma \gamma_{\delta \beta} + \partial_\beta \gamma_{\delta \gamma} - \partial_\delta \gamma_{\beta \gamma}),
\end{eqnarray}
where $\lambda^\alpha_{\beta \gamma}$ are the 3-dimensional Christoffel symbols formed using the metric $\gamma_{\alpha\beta}$.
The three-dimensional Ricci tensor $P_{\alpha \beta}$ associated with $\gamma_{\alpha \beta}$ is then,
\begin{equation}
P_{\alpha \beta} = \partial_\mu \Gamma^\mu_{\alpha \beta} - \partial_\beta \Gamma^\mu_{\alpha \mu} +
\Gamma^\mu_{\alpha \beta}\Gamma^\epsilon_{\mu \epsilon} - \Gamma^\mu_{\alpha \epsilon}\Gamma^\epsilon_{\beta \mu}.
\label{eqP_ab}
\end{equation}
It is very useful to obtain asymptotic expansions  for the spatial Ricci tensor $P_{\alpha \beta}$ and its trace valid near the sudden singularity. This is done below in several steps, and should be thought of as a necessary step to highlight the behaviour of the various components of the 4-dimensional Ricci curvature introduced consequently. First, we introduce the symbols,
\begin{eqnarray}
A_{\alpha \beta \epsilon} &=& \partial_\beta a_{\alpha \epsilon} + \partial_\alpha a_{\beta \epsilon} - \partial_\epsilon a_{\alpha \beta}, \\
B_{\alpha \beta \epsilon} &=& \partial_\beta b_{\alpha \epsilon} + \partial_\alpha b_{\beta \epsilon} - \partial_\epsilon b_{\alpha \beta}, \\
C_{\alpha \beta \epsilon} &=& \partial_\beta c_{\alpha \epsilon} + \partial_\alpha c_{\beta \epsilon} - \partial_\epsilon c_{\alpha \beta},
\end{eqnarray}
and using the basic metric expansion around the sudden singularity, Eq. (\ref{seriesgab}), we find that,
\begin{eqnarray}
\Gamma^\mu_{\alpha \beta} &=& \lambda^\mu_{\alpha \beta} = \frac{1}{2}a^{\mu \epsilon}A_{\alpha \beta \epsilon} + \frac{1}{2}(a^{\mu \epsilon} B_{\alpha \beta \epsilon} - b^{\mu \epsilon} A_{\alpha \beta \epsilon})t + \frac{1}{2}(a^{\mu \epsilon} C_{\alpha \beta \epsilon} - c^{\mu \epsilon} A_{\alpha \beta \epsilon})t^n + O(t^2) \nonumber \\
&=& (\lambda^\mu_{\alpha \beta})_0 + (\lambda^\mu_{\alpha \beta})_1 t + (\lambda^\mu_{\alpha \beta})_n t^n + O(t^2),
\label{seriesG}
\end{eqnarray}
with the coefficients $(\lambda^\mu_{\alpha \beta})_i$ obviously defined.
Then Eq. (\ref{eqP_ab}) implies,
\begin{equation}
P_{\alpha \beta} = (P_{\alpha \beta})_0 + (P_{\alpha \beta})_1 t + (P_{\alpha \beta})_n t^n + O(t^2),
\label{seriesPab}
\end{equation}
where the corresponding coefficients of the first few orders for the spatial Ricci 3-curvature are expressed using the $(\lambda^\mu_{\alpha \beta})_i$'s as follows,
\begin{eqnarray}
(P_{\alpha \beta})_0 &=& \partial_\mu (\lambda^\mu_{\alpha \beta})_0 -
\partial_\beta (\lambda^\mu_{\alpha \mu})_0 +
(\lambda^\mu_{\alpha \beta})_0 (\lambda^\epsilon_{\mu \epsilon})_0 -
(\lambda^\mu_{\alpha \epsilon})_0 (\lambda^\epsilon_{\beta \mu})_0, \\
(P_{\alpha \beta})_1 &=& \partial_\mu (\lambda^\mu_{\alpha \beta})_1 - \partial_\beta (\lambda^\mu_{\alpha \mu})_1 +
(\lambda^\mu_{\alpha \beta})_0 (\lambda^\epsilon_{\mu \epsilon})_1 +
(\lambda^\epsilon_{\mu \epsilon})_0 (\lambda^\mu_{\alpha \beta})_1 - \nonumber \\
&&(\lambda^\mu_{\alpha \epsilon})_0 (\lambda^\epsilon_{\beta \mu})_1 -
(\lambda^\epsilon_{\beta \mu})_0 (\lambda^\mu_{\alpha \epsilon})_1, \\
(P_{\alpha \beta})_n &=& \partial_\mu (\lambda^\mu_{\alpha \beta})_n - \partial_\beta (\lambda^\mu_{\alpha \mu})_n +
(\lambda^\mu_{\alpha \beta})_0 (\lambda^\epsilon_{\mu \epsilon})_n +
(\lambda^\epsilon_{\mu \epsilon})_0 (\lambda^\mu_{\alpha \beta})_n - \nonumber \\
&&(\lambda^\mu_{\alpha \epsilon})_0 (\lambda^\epsilon_{\beta \mu})_n -
(\lambda^\epsilon_{\beta \mu})_0 (\lambda^\mu_{\alpha \epsilon})_n.
\end{eqnarray}
Then the mixed components of the Ricci 3-curvature are given by,
\begin{eqnarray}
P^\beta_\alpha &=& \gamma^{\beta \mu} P_{\mu \alpha} \nonumber \\
&=& a^{\beta \mu}(P_{\mu \alpha})_0 + [a^{\beta \mu}(P_{\mu \alpha})_1 - b^{\beta \mu}(P_{\mu \alpha})_0] t +
[a^{\beta \mu}(P_{\mu \alpha})_n - c^{\beta \mu}(P_{\mu \alpha})_0] t^n + O(t^2) \nonumber \\
&=& (P^\beta_\alpha)_0 + (P^\beta_\alpha)_1 t + (P^\beta_\alpha)_n t^n + O(t^2),
\label{seriesPabmixed}
\end{eqnarray}
which implies that the trace expansion for the $\textrm{Tr}P^\alpha_\beta$ is given by,
\begin{eqnarray}
P &=& P^\alpha_\alpha =  \delta^\alpha_\beta P^\beta_\alpha \nonumber \\
&=& a^{\alpha \mu}(P_{\mu \alpha})_0 + [a^{\alpha \mu}(P_{\mu \alpha})_1 - b^{\alpha \mu}(P_{\mu \alpha})_0] t +
[a^{\alpha \mu}(P_{\mu \alpha})_n - c^{\alpha \mu}(P_{\mu \alpha})_0] t^n + O(t^2) \nonumber \\
&=& P_0 + P_1 t + P_n t^n + O(t^2).
\label{seriesP}
\end{eqnarray}
The zeroth-order term is then given explicitly by the expansion,
\begin{eqnarray}
P_0 &=& a^{\alpha \beta}(P_{\alpha \beta})_0 \nonumber \\
&=& a^{\alpha \beta}[\partial_\mu \widetilde{\Gamma}^\mu_{\alpha \beta} - \partial_\beta \widetilde{\Gamma}^\mu_{\alpha \mu} +
\widetilde{\Gamma}^\mu_{\alpha \beta}\widetilde{\Gamma}^\epsilon_{\mu \epsilon} - \widetilde{\Gamma}^\mu_{\alpha \epsilon}\widetilde{\Gamma}^\epsilon_{\beta \mu}] \nonumber \\
&=& \frac{1}{2}a^{\alpha \beta}[\partial_\mu (a^{\mu \epsilon}A_{\alpha \beta \epsilon}) -
\partial_\beta (a^{\mu \epsilon}A_{\alpha \mu \epsilon}) +
\frac{1}{2}a^{\mu \gamma}a^{\epsilon \delta}A_{\alpha \beta \gamma}A_{\mu \epsilon \delta} -
\frac{1}{2}a^{\mu \gamma}a^{\epsilon \delta}A_{\alpha \epsilon \gamma}A_{\beta \mu \delta}] \nonumber \\
&=& \frac{1}{2}a^{\alpha \beta} \big{\{} \partial_\mu [(a^{\mu \epsilon} (\partial_\beta a_{\alpha \epsilon} + \partial_\alpha a_{\beta \epsilon} - \partial_\epsilon a_{\alpha \beta})]  - \partial_\beta [(a^{\mu \epsilon} (\partial_\mu a_{\alpha \epsilon} + \partial_\alpha a_{\mu \epsilon} - \partial_\epsilon a_{\alpha \mu})] \nonumber + \\
&& \frac{1}{2}a^{\mu \gamma}a^{\epsilon \delta}(\partial_\beta a_{\alpha \gamma} + \partial_\alpha a_{\beta \gamma} - \partial_\gamma a_{\alpha \beta})(\partial_\epsilon a_{\mu \delta} + \partial_\mu a_{\epsilon \delta} - \partial_\delta a_{\mu \epsilon}) \nonumber - \\
&& \frac{1}{2}a^{\mu \gamma}a^{\epsilon \delta}(\partial_\epsilon a_{\alpha \gamma} + \partial_\alpha a_{\epsilon \gamma} - \partial_\gamma a_{\alpha \epsilon})(\partial_\mu a_{\beta \delta} + \partial_\beta a_{\mu \delta} - \partial_\delta a_{\beta \mu}) \big{\}}.
\label{eqP0}
\end{eqnarray}

We can now compute the various expansions  of the Ricci tensor near the sudden singularity using the above results and the standard formulae for the Ricci tensor decomposition in a synchronous frame \cite{LL}.
\begin{eqnarray}
R_{0}^{0} &=& -\frac{1}{2}\partial_t K - \frac{1}{4}K^\beta_\alpha K^\alpha_\beta \\
&=& -\frac{1}{2}n(n-1)c\;t^{n-2} + (\frac{1}{4}b^\beta_\alpha b^\alpha_\beta - d) + \frac{1}{2}n^2 b^\beta_\alpha c^\alpha_\beta t^{n-1} + \cdots ,\label{seriesR00} \\
R_{\alpha }^{0} &=& \frac{1}{2}(\nabla_\beta K^\beta_\alpha - \nabla_\alpha K) \\
&=& \frac{1}{2}(\nabla_{\beta}b_{\alpha}^{\beta }-\nabla_{\alpha}b)+
\frac{1}{2}n(\nabla_{\beta}c_{\alpha}^{\beta }-\nabla_{\alpha}c)\;t^{n-1} + \\
&& [(\nabla_\beta d^\beta_\alpha - \nabla_\alpha d) - \frac{1}{2}\nabla_\beta (b^\beta_\gamma b^\gamma_\alpha) +
\frac{1}{2}\nabla_\alpha (b^\beta_\gamma b^\gamma_\beta)]t + \cdots , \\
R^{\beta }_{\alpha } &=& -P^\beta_\alpha - \frac{1}{2}\partial_t K^\beta_\alpha - \frac{1}{4}KK^\beta_\alpha \\
&=& -\frac{1}{2}n(n-1)c^{\beta }_{\alpha }\;t^{n-2} +
[-a^{\beta \gamma}(P_{\gamma \alpha})_0 - \frac{1}{4}bb^\beta_\alpha + \frac{1}{2}b^\beta_\gamma b^\gamma_\alpha - d^\beta_\alpha] - \\
&& \frac{1}{4}n(bc^\beta_\alpha + cb^\beta_\alpha -2n b^\beta_\gamma c^\gamma_\alpha - 2b^\gamma_\alpha c^\beta_\gamma)t^{n-1} + \cdots .\label{rab}
\end{eqnarray}
Consequently, for the scalar curvature we obtain the expansion,
\begin{equation}\label{seriesR}
R = -n(n-1)ct^{n-2} + (-P_0 - \frac{1}{4}b^2 + \frac{3}{4}b^\beta_\alpha b^\alpha_\beta - 2d) +
\frac{1}{2}n[(2n+1)b^\beta_\alpha c^\alpha_\beta - bc]t^{n-1} + \cdots.
\end{equation}

\section{The Brans-Dicke equations at the sudden singularity}
In this Section we find the asymptotic nature of the various components of the Brans-Dicke system of equations (\ref{bd})-(\ref{conlaws}) as we approach the sudden future singularity. For this purpose, we shall first split the various terms in the equations in the synchronous system and then  in the second subsection find the balancing conditions to the corresponding leading orders. As a result, we shall discover the asymptotic  expansions for $\rho, p$ and $u_\alpha$  near the sudden singularity,  and set asymptotic constraints as functional relations for the consistency of the whole scheme.

\subsection{Splittings}
We assume that the stress energy tensor has the form of a perfect fluid,
\begin{equation}
T_{j}^{i}=  (\rho +p)u^{i}u_{j}-p\delta _{j}^{i},
\end{equation}%
with the unit 4-velocity $u^{i}=(u^{0},u^{\alpha })$ with $u^{0}=u_{0}$, and $u_{i}u^{i}=1$, so that,
\be
u_{0}^{2}=1+u_\alpha u^\alpha,
\ee
which means that the three arbitrary components $u^\alpha$ of the velocity vector field determine $u^0$.  It follows that near the sudden singularity,
\begin{equation}
u_{0}^{2}=1+(a_{\alpha \beta }+b_{\alpha \beta }\;t+c_{\alpha
\beta }\;t^{n} + \cdots )u^{\alpha }u^{\beta }.
\label{4vel}
\end{equation}
Since the simplest choice for the for the velocity vector field $u^i$ tangent to the streamlines and for the vector field $n^i$ normal to the hypersurface $\Sigma$ is to set (\cite{LL}, Section 97),
\be
u^i=n^i=(1,0),
\ee
we assume that in the general case of an arbitrary unit velocity vector field, to leading order we have,
\be
u^0\sim 1, \quad u^\alpha\sim 0.
\ee
Hence, to leading order we find,
\begin{eqnarray}
T_{0}^{0} &=&(\rho +p)u^0u_{0}-p\sim\rho ,
\label{T00} \\
T_{\alpha }^{0} &=&(\rho +p)u^{0}u_{\alpha }\sim(\rho +p)u_{\alpha },
\label{T0a} \\
T^{\beta }_{\alpha } &=&(\rho +p)u^{\beta }u_{\alpha }-p\delta ^{\beta}_{\alpha }\sim -p\delta ^{\beta }_{\alpha }.
\label{Tab}
\end{eqnarray}
The conservation laws $\nabla_i T^i_j=0$ for the matter content split into temporal and spatial components as follows:
\begin{equation}
\partial_t \rho + \partial_\alpha [(\rho + p)u^\alpha] + \frac{1}{2}K(\rho + p) +
\lambda^\alpha_{\alpha \beta}(\rho + p)u^\beta = 0,
\label{conlaw0}
\end{equation}
\begin{equation}
\partial_t[(\rho + p)u_\alpha] - \partial_\alpha p + \frac{1}{2}K (\rho + p)u_\alpha = 0,
\label{conlawalpha}
\end{equation}%
and we are faced with the problem of finding the leading orders of the various terms in these equations, and similarly in the equation of motion of the scalar field, and also in the full Brans-Dicke equations (see below).

For the various derivatives of the scalar field $\phi=\phi(t, x^\alpha)$, we find the results:
\begin{eqnarray}
\partial_t \phi &=& \phi_1 + n\phi_n t^{n-1} + O(t) \\
{\partial_t}^2 \phi &=& n(n-1)\phi_n t^{n-2} + O(1) \\
\partial_\alpha \phi &=& (\partial_\alpha \phi_0) + (\partial_\alpha \phi_1)t + (\partial_\alpha \phi_n)t^n + O(t^2) \\
\partial^\beta \phi \partial_\alpha \phi &=& -\gamma^{\beta \gamma}\partial_\gamma \phi \partial_\alpha \phi=-\alpha^{\beta \gamma}(\partial_\gamma \phi_0) (\partial_\alpha \phi_0)+\nonumber\\ &+&
[-\alpha^{\beta \gamma}(\partial_\gamma \phi_0) (\partial_\alpha \phi_1) -
\alpha^{\beta \gamma}(\partial_\gamma \phi_1) (\partial_\alpha \phi_0) +
b^{\beta \gamma}(\partial_\gamma \phi_0) (\partial_\alpha \phi_0)]t\nonumber \\ &+&
[-\alpha^{\beta \gamma}(\partial_\gamma \phi_0) (\partial_\alpha \phi_n) -
\alpha^{\beta \gamma}(\partial_\gamma \phi_n) (\partial_\alpha \phi_0) -
c^{\beta \gamma}(\partial_\gamma \phi_0) (\partial_\alpha \phi_0)]t^n \nonumber\\&+& O(t^2) \\
\nabla^0 \nabla_0 \phi &=& {\partial_t}^2 \phi = n(n-1)\phi_n t^{n-2} + O(1) \\
\nabla^0 \nabla_\alpha \phi &=& \partial_t (\partial_\alpha \phi) - \frac{1}{2}K^\beta_\alpha \partial_\beta \phi \nonumber \\
&=& [(\partial_\alpha \phi_1) - \frac{1}{2}b^\beta_\alpha (\partial_\beta \phi_0)] +
n[(\partial_\alpha \phi_n) - \frac{1}{2}c^\beta_\alpha (\partial_\beta \phi_0)]t^{n-1} + O(t) \\
\nabla^\beta \nabla_\alpha \phi &=& -\gamma^{\beta \gamma}\partial_\gamma (\partial_\alpha \phi) +
\frac{1}{2}K^\beta_\alpha \partial_t \phi + \gamma^{\beta \gamma}\lambda^\delta_{\gamma \alpha}\partial_\delta \phi \nonumber \\
&=& [-\alpha^{\beta \gamma}\partial_\gamma (\partial_\alpha \phi_0) + \frac{1}{2} \phi_1 b^\beta _\alpha
+ \alpha^{\beta \gamma}(\lambda^\delta_{\gamma \alpha})_0 (\partial_\delta \phi_0)] +\nonumber\\ &+&
\frac{1}{2}n(\phi_n b^\beta _\alpha + \phi_1 c^\beta _\alpha)t^{n-1} + O(t) \\
\Box_g \phi &=& {\partial_t}^2 \phi + \frac{1}{2}K\partial_t \phi - \gamma^{\alpha \beta}\partial_\beta (\partial_\alpha \phi) + \gamma^{\alpha \beta}\lambda^\gamma_{\beta \alpha}\partial_\gamma \phi \nonumber \\
&=& n(n-1)\phi_n t^{n-2} + O(1).
\end{eqnarray}
We have the equation of motion for the scalar field, Eq. (\ref{Boxphi}), namely,
\begin{equation}
(3 + 2\omega)\Box \phi = 8\pi(\rho - 3p),
\label{Boxphi2}
\end{equation}
Lastly, using the Brans-Dicke  equations (\ref{bd}),
\begin{equation}
\phi^2 (R_{j}^{i}-\frac{1}{2}\delta^i_j R)=8\pi \phi T_{j}^{i} + \omega (\partial^i \phi \partial_j \phi - \frac{1}{2}\delta^i_j \partial^k \phi \partial_k \phi) + \phi(\nabla^i \nabla_j \phi - \delta^i_j \Box_g \phi),
\label{BDeq}
\end{equation}%
split into the $\binom{0}{0}$ component,
\begin{equation}
\phi^2 (R^0_0 - \frac{1}{2}R) = 8\pi \phi \rho + \frac{1}{2}\omega[(\partial_t \phi)^2 -
\partial^\alpha \phi \partial _\alpha \phi] - \phi \nabla^\alpha \nabla_\alpha \phi,
\label{BD00}
\end{equation}
the $\binom{0}{\alpha }$ components,
\begin{equation}
\phi^2 R^0_\alpha = 8\pi \phi (\rho + p)u_\alpha + \omega \partial_t \phi \partial_\alpha \phi +
\phi \nabla^0 \nabla_\alpha \phi,
\label{BD0a}
\end{equation}%
and the $\binom{\beta }{\alpha }$ components,
\begin{equation}
\phi^2 (R^{\beta }_{\alpha} - \frac{1}{2}\delta^\beta_\alpha R) = -8\pi \phi p \delta^\beta_\alpha +
\omega[\partial^\beta \phi \partial_\alpha \phi - \frac{1}{2}\delta^\beta_\alpha(\partial_t \phi)^2 -
\frac{1}{2}\delta^\beta_\alpha \partial^\gamma \phi \partial_\gamma \phi] +
\phi(\nabla^\beta \nabla_\alpha \phi - \delta^\beta_\alpha \Box \phi).
\label{BDab}
\end{equation}

\subsection{Balancing}
We start with the $\binom{0}{0}$ component in order to evaluate the energy density and then the
trace of the $\binom{\beta }{\alpha }$ equations to calculate the pressure.
Based on the resulting equations of the the energy density and the pressure, the 4-velocity can be found from the $\binom{0}{\alpha }$ components. We further take into account the $\binom{\beta }{\alpha }$ components to get
restrictions on the arbitrary functions of $a_{\alpha \beta },b_{\alpha \beta },$ and $c_{\alpha \beta}$.

From the $\binom{0}{0}$ component, given by (\ref{BD00}), we obtain the relation of the energy density, which is,
\begin{eqnarray}
8\pi \rho &=&  \phi_0 \left(\frac{1}{2}P_0 + \frac{1}{8}b^2 - \frac{1}{8} b^\beta_\alpha b^\alpha_\beta\right) -
\frac{1}{2}\omega \left[\frac{(\phi_1)^2}{\phi_0} + a^{\alpha \beta}\frac{(\partial_\alpha \phi_0)(\partial_\beta \phi_0)}{\phi_0}\right]
\nonumber \\
&+& \left[-a^{\alpha \beta}\partial_\beta (\partial_\alpha \phi_0) +\frac{1}{2}\phi_1 b
+ a^{\alpha \beta}(\lambda^\gamma_{\alpha \beta})_0 (\partial_\gamma \phi_0)\right]  \nonumber \\
&+& \left[\frac{1}{4}n\phi_0 (bc - b^\beta_\alpha c^\alpha_\beta) - \omega n \frac{\phi_1 \phi_n}{\phi_0}
+ \frac{1}{2}n(\phi_n b + \phi_1 c)\right]t^{n-1} + O(t) ,
\label{seriesdensity}
\end{eqnarray}%
where $P_0$ is the zeroth-order term of the Ricci scalar associated with $a_{\alpha \beta}$. This is an expansion of the form,
\be
\rho=\rho_0+\rho_{n-1} t^{n-1}+\rho_1t+\cdots,
\ee
with the coefficients $\rho_i$ representing the shown functions which are independent of the time.

The trace of the $\binom{\beta}{\alpha}$ component, given by (\ref{BDab}), implies,
\begin{equation}
\phi^2 (R^\alpha_\alpha - \frac{3}{2}R) = -24\pi \phi p -\frac{1}{2} \omega[3(\partial_t \phi)^2 +
\partial^\alpha \phi \partial_\alpha \phi] - \phi(3\nabla^0 \nabla_0 \phi + 2\nabla^\alpha \nabla_\alpha \phi),
\label{BDaa}
\end{equation}
from which, for the terms of $O(t^{n-2})$, we find,
\begin{equation}
(\phi_0)^2 \left[-\frac{1}{2}n(n-1)c+\frac{3}{2}n(n-1)c\right]=-24\pi \phi_0 p_{n-2}-3n(n-1)\phi_0 \phi_n.
\label{seriespressure_n-2}
\end{equation}
Thus we end up with the expansion of the pressure, namely,
\begin{equation}
8\pi p = -\frac{1}{3}n(n-1)(\phi_0 c + 3\phi_n)t^{n-2} + O(1),
\label{seriespressure}
\end{equation}%
which we write generally as,
\be
p=p_{n-2}t^{n-2}+p_0+p_{n-1} t^{n-1}+\cdots,
\ee
where again the coefficients are time-independent functions.

We note, however, that the $\binom{\alpha }{\beta }$ components of the Brans-Dicke equations for $\alpha\neq\beta$, give for the $(n-2)$-order terms,
\begin{equation}
-\frac{1}{2}n(n-1)(\phi_0)^2(c^\beta_\alpha - c\delta^\beta_\alpha) =
\frac{1}{3}n(n-1)\phi_0(\phi_0 c + 3\phi_n)\delta^\beta_\alpha - n(n-1)\phi_0 \phi_n \delta^\beta_\alpha
\label{eqc_initial}
\end{equation}%
which after simplifying gives,
\begin{equation}
c^{\beta }_{\alpha }=\frac{c}{3}\delta ^{\beta }_{\alpha}.
\label{eqc}
\end{equation}%
Equations (\ref{eqc}) represent six functional constraining relations between the initial data $c_{\alpha \beta}$ from which leave only one out of the six components of $c_{\alpha \beta}$ arbitrary.

Additionally, based on the dominant term of the $\binom{0}{\alpha }$ components given by (\ref{BD0a}), we find,
\begin{equation}
8\pi \phi_0 p_{n-2} u_{\alpha }^{2-n} = \frac{1}{2}(\phi_0)^2(\nabla_\beta b^\beta_\alpha - \nabla_\alpha b) - \omega \phi_1 (\partial_\alpha \phi_0) - \phi_0 (\partial_\alpha \phi_1) +
\frac{1}{2}\phi_0 b^\beta_\alpha (\partial_\beta \phi_0),
\label{velocities_initial}
\end{equation}
that is the three relations for the arbitrary  spatial components of velocity vector field are,
\bq
u_{\alpha}&=&\frac{3}{n(n-1)(\phi_0 c + 3\phi_n)}\left[-\frac{1}{2}(\phi_0)^2(\nabla_\beta b^\beta_\alpha - \nabla_\alpha b) + \omega \phi_1 (\partial_\alpha \phi_0) + \phi_0 (\partial_\alpha \phi_1) -\right.\nonumber\\
&-&\left.
\frac{1}{2}\phi_0 b^\beta_\alpha (\partial_\beta \phi_0)\right]t^{2-n}.
\label{velocities}
\eq
This is an expansion of the form,
\be
u_{\alpha}=(u_{\alpha})_{2-n}t^{2-n}+(u_{\alpha})_1 t+\cdots .
\ee
We note that the combination $\phi_0 c + 3\phi_n$ that appears in the denominator of the Eq. (\ref{velocities}) is not zero due to the equation for the pressure (\ref{seriespressure}).

Using the results for the energy density and the pressure, the equation of motion of the scalar field, Eq.  (\ref{Boxphi2}), gives,
\begin{equation}
(3+2\omega)n(n-1)\phi_n = -3[-\frac{1}{3}n(n-1)\phi_0 c - n(n-1)\phi_n],
\label{eqphi_n_initial}
\end{equation}
namely, we arrive at the functional relation,
\begin{equation}
\phi_n = \frac{\phi_0 c}{2\omega},
\label{eqphi_n}
\end{equation}
which represents one connection between the data in the  $\phi$ expansion.

Further, from the analysis of the asymptotics of the conservation equations, no new constraining relation can be found.  This is seen as follows. From the $(n-2)$-order term of the time component of the conservations equations, Eq. (\ref{conlaw0}), we find,
\bq
&&(n-1)\left[\frac{1}{4}n\phi_0 (bc-b^\beta_\alpha c^\alpha_\beta)-\omega n\frac{\phi_1 \phi_n}{\phi_0} + \frac{1}{2}n(\phi_n b + \phi_1 c)\right]-\nonumber\\
&&-\frac{1}{2}b\left[\frac{1}{3}n(n-1)\phi_0 c+n(n-1)\phi_n\right]=0,
\label{conlaw0_initial}
\eq
from which we obtain,
\begin{equation}
\phi_1 \phi_n =\frac{\phi_0 \phi_1 c}{2\omega}.
\label{conlaw0_initial2}
\end{equation}%
However this last equation is not a new one as it follows directly from Eq. (\ref{eqphi_n}).

Finally, for the $(n-2)-$order, the spatial part of the stress-energy conservation (\ref{conlawalpha}) is expressed as,
\begin{eqnarray}
&&\frac{1}{2}n(n-1)(\phi_0)^2 (\nabla_\beta c^\beta_\alpha - \nabla_\alpha c) - \omega n(n-1)\phi_n (\partial_\alpha \phi_0) -\nonumber\\ &-&
n(n-1)\phi_0 [(\partial_\alpha \phi_n) - \frac{1}{2}c^\beta_\alpha (\partial_\beta \phi_0)] \nonumber \\
&=& -\phi_0 [\frac{1}{3}n(n-1)(\partial_\alpha \phi_0)c + \frac{1}{3}n(n-1)\phi_0(\partial_\alpha c) +
n(n-1)(\partial_\alpha \phi_n)].
\label{conlawalpha_initial}
\end{eqnarray}%
Using (\ref{eqphi_n}), we find,
\begin{equation}
\frac{1}{2}(\phi_0)^2 \nabla_\beta c^\beta_\alpha - \frac{1}{6} (\phi_0)^2 \nabla_\alpha c = 0,
\label{conlawalpha_initial2}
\end{equation}%
which  follows from the constraint relation (\ref{eqc}), and so is not new.

\section{Counting}
In general, we expect there will be
$6\times g_{\alpha \beta}$ and $6\times \dot{g}_{_{\alpha\beta }}$, plus $3$ free velocity components $u_{\alpha },$ plus $2$ from the pressure $p$ and the density $\rho $, and $2$ additional from $\phi$ and $\dot{\phi}$ giving a total of $19$ independent functions.
We can remove four of these by using the $G^0_0$ and $G_{a}^{0}$ constraints and four more by using the
general coordinate covariances. This leaves a total of $11$ free functions
expected in the general solution for the metric locally. If an equation of
state had been assumed to relate the pressure to the density this number
would have been reduced by $1$ to $10$.

Hence taking into account relations (\ref{seriesdensity}), (\ref{seriespressure}), (\ref{eqc}), (\ref{velocities}) and (\ref{eqphi_n}), we have $6+6+1=13$ independent functions from the initial data $(a_{\alpha \beta}, b_{\alpha \beta}, c_{\alpha \beta})$, plus $1+1+0=2$ independent functions from $(\phi_0, \phi_1, \phi_n)$. Therefore, in total we have found $15$ independent functions. Subtracting the $4$ coordinate covariances which may still be used to remove four functions, leaves $11$
independent arbitrary functions of the three space coordinates on a surface of constant $t$ time.

This is the maximal number of independent arbitrary spatial functions expected in a local representation of part of the general solution of Brans-Dicke's equations near a sudden singularity.

\section{Sudden singularities as shock waves}
The main result in this work that the asymptotic approach to the sudden singularity in Brans-Dicke theory is part of the general solution of the theory allows us to make some comments as to the general character of the sudden singularity in the present context. This problem was addressed for general relativity in Ref. \cite{bc13}, where it was shown that the general approach to the inhomogeneous and anisotropic sudden singularity is geodesically complete. The proof  was based on an analysis of the solutions of the geodesic equations having a $\mathcal{C}^2$ character near the sudden singularity. Here we shall present a new proof of the completeness of geodesics, and also extend it to the context of Brans-Dicke theory.

Before we proceed, we note that  in both general relativity and Brans-Dicke theories the asymptotic behaviours of the various components of the Ricci tensor on approach to the sudden singularity are identical, namely, $R_{00}\sim t^{n-2},R_{0\alpha }\sim t^{0},R_{\alpha \gamma }\sim t^{n-2},$
while $u^{0}\sim t^{0},u^{\alpha }\sim t^{2}$. Hence, following the same analysis as in Ref. \cite{bc13}, we arrive at the conclusion that  the Brans-Dicke sudden singularity is also weak in the senses of Tipler \cite{tip} and Krolak  \cite{kr}.

We now need to estimate the combination $R_{ij}u^{i}u^{j}$ asymptotically on approach to the sudden singularity. We start with general relativity. Using Eq. (\ref{eqc}), to express the symbol $c^{\alpha}_{\beta}$ in the leading term of $R^{\alpha}_{\beta}$ in Eq. (\ref{rab}), we find that to leading order,
\be \label{tcc}
R_{ij}u^{i}u^{j}\sim -c\,t^{n-2}.
\ee
Therefore, the strong energy condition, which for the Einstein equations implies that $R_{ij}u^{i}u^{j}\geq 0$, gives,
\be
c<0.
\ee
Therefore $R_{ij}u^{i}u^{j}$ does not change sign  and remains positive during the approach to the sudden singularity if we assume that $c<0$ initially. If we imagine a geodesic congruence starting at some earlier time and approaching the sudden future singularity, then asymptotically the expansion of the congruence is given by Eq. (\ref{k}), namely,
\be\label{k1}
K\sim b+nct^{n-1},
\ee
and the change in $K$ is given by Eq. (\ref{dotk}), that is to leading order,
\be\label{k2}
\dot{K}\sim n(n-1)ct^{n-2}.
\ee
Therefore even if we assume that initially (that is before the sudden singularity) the congruence is converging, that is,
\be
b<0,
\ee
so that the right hand side in (\ref{k1}) is negative, and further that the strong energy condition is satisfied, namely $c<0$, then we find,
\be \label{focgr}
K\rightarrow b,\quad\dot{K}\rightarrow -\infty,\quad as\quad t\rightarrow 0.
\ee
Here $K$ tends to the constant $b$ in the finite proper time remaining to the sudden singularity, instead of $-\infty$ as one would expect from the focussing theorem of the standard singularity theorems. The fact that $K$ cannot diverge to $-\infty$ there means that the sudden singularity can never be a point conjugate to \emph{any} earlier point in spacetime.

The reason for this difference can be seen most clearly by looking at the balancing of the various terms in the Landau-Raychaudhuri equation,
\be \label{ray}
\dot{K}=-\frac{1}{3}K^2-\sigma^{ij}\sigma_{ij}-R_{ij}u^{i}u^{j}.
\ee
Because of the fact that the sudden singularity metric contains the crucial term $t^n,n\in(1,2)$, the term $\dot{K}$ does not balance with the term $K^2$ (which is subdominant from Eq. (\ref{k1})) as in the case of the singularity theorems, but here it balances with the last term. Therefore the evolution of $K$ and $\dot{K}$ is separate in the sudden singularity  case presently, because these two quantities are not of the same asymptotic order.

This result has an interesting interpretation if we take into account the fact that the expansion $K$ represents the fractional rate of change of the cross-sectional volume of the geodesic congruence. At the sudden singularity  there is a minimum volume, given by $b$, for geodesics to pass without intersecting each other,  and  they cannot converge further to a region smaller than this there. Therefore they escape to the future leaving behind, at the sudden singularity, a hypersurface of finite volume which has some discontinuity in the second derivatives of the metric, much like a shock wave.

Let us finally take the above argument to see how it changes in the present context of Brans-Dicke theory. The whole argument above basically remains the same for the Brans-Dicke case, because the last term in the Landau-Raychaudhuri equation is of the same leading order as in the case of general relativity. The only difference is in the strong energy condition for the matter present in the Brans-Dicke equations, from which we get an extra condition. The strong energy condition dictates that,
\be
(T_{ij}-\frac{1}{2}\delta^i_j T)u^{i}u^{j}\geq 0,
\ee
which is our case translates to the condition,
\be
\left(1+\frac{3}{2\omega}\right)\phi_0 c\leq 0,
\ee
taking into account the asymptotic functional relation (\ref{eqphi_n}). When the term in the brackets is positive, then for the previous argument to continue to be valid here, we need only to further assume that $\phi_0>0$. Then the strong energy condition implies that $c<0$, and the timelike convergence condition follows from the asymptotic relation (\ref{tcc}).
\section{Discussion}
Our foregoing results  indicate that near the sudden singularity the Brans-Dicke equations with a fluid source admit a solution with 11 arbitrary functions as it is necessary for a general solution in that theory. This result compares with the corresponding one in general relativity, where the solution there has 9 arbitrary functions and was also a general one \cite{bct}.

In general relativity, near a non-sudden region containing a spacetime singularity, a similar situation arises in the non-singular approach to a quasi-isotropic de Sitter spacetime in the presence of a positive cosmological constant as in Refs. \cite{no hair}-\cite{rend}, or with an ultrastiff fluid with $p>\rho$ as in \cite{hs}. The only other result known with comparable simplicity is the vacuum, non-singular solution in $R+\epsilon R^2$ gravity constructed in \cite{ct16} and containing 16 arbitrary data.

As we showed, the generic sudden singularity region in Brans-Dicke theory contains no geodesic incompleteness and has the character of a shock wave. We would expect that these results will continue to hold in general $f(R)$ theory near a sudden singularity region and other modified gravity models.
\section*{Acknowledgments}
\noindent Between the original submission of this paper to the arXiv and its submission to this Journal, John Barrow passed away. He was a great scientist, mentor, and friend, and his fond memory will always remain in our hearts. The current version of this paper is essentially the same as the one originally announced in the arXiv last September.

\appendix
\section{Appendix: The homogeneous scalar field}
We now consider the function counting problem in the special case where the metric expansion is as before,
\begin{equation}
\gamma _{\alpha \beta }=a_{_{\alpha \beta }}+b_{_{\alpha \beta
}}t+c_{_{\alpha \beta }}t^{n}+ O(t^2),
\label{seriesgabhom}
\end{equation}%
but the scalar field expansion near the sudden singularity is `homogeneous', meaning,
\begin{equation}
\phi = \phi_0 + \phi_1 t + \phi_n t^{n} + O(t^2),
\label{seriesphihom}
\end{equation}%
with $1<n<2$, but the coefficients $\phi_i $ are constants. The reason we need to demonstrate this special case in full is that when one drops the spatial gradients of the scalar field completely from all equations, one cannot be sure that no functional constraint is lost or changed  in the process, especially those constraints associated with the conservation laws. Of course, if one first works out the problem in the homogeneous scalar field case, it is not possible to guess the results that hold for the inhomogeneous case as in the main body of this paper.

The inverse metric tensor is given as before,
\begin{equation}
\gamma ^{\alpha \beta }=a^{\alpha \beta }-b^{\alpha \beta }\;t-c^{\alpha
\beta }\;t^{n} + O(t^2)
\label{invgabhom},
\end{equation}%
and the series related with the extrinsic curvature, its derivatives, and its contractions are the same. The components of the Ricci tensor have also the same forms (\ref{seriesR00}) - (\ref{seriesR}). However, for the derivatives of the scalar field with  $\phi=\phi(t)$, we find,
\begin{eqnarray}
\partial_t \phi &=& \phi_1 + n\phi_n t^{n-1} + O(t) \\
{\partial_t}^2 \phi &=& n(n-1)\phi_n t^{n-2} + O(1) \\
\nabla^\beta \nabla_\alpha \phi &=& \frac{1}{2}K^\beta_\alpha \partial_t \phi \nonumber \\
&=& \frac{1}{2} \phi_1 b^\beta _\alpha + \frac{1}{2}n(\phi_n b^\beta _\alpha + \phi_1 c^\beta _\alpha)t^{n-1} + O(t)  \\
\Box_g \phi &=& {\partial_t}^2 \phi + \frac{1}{2}K\partial_t \phi \nonumber \\
&=& n(n-1)\phi_n t^{n-2} + O(1).
\end{eqnarray}
From the Brans-Dicke equations
\begin{equation}
\phi^2 (R_{j}^{i}-\frac{1}{2}\delta^i_j R)=8\pi \phi T_{j}^{i} + \omega (\partial^i \phi \partial_j \phi - \frac{1}{2}\delta^i_j \partial^k \phi \partial_k \phi) + \phi(\nabla^i \nabla_j \phi - \delta^i_j \Box_g \phi),
\label{BDeqhom}
\end{equation}%
we use the $\binom{0}{0}$ component as before to calculate the energy density and the
trace of the $\binom{\beta }{\alpha }$ equations to calculate the pressure.
The 4-velocity can be calculated then from the $\binom{0}{\alpha }$
components. Also we use the $\binom{\beta }{\alpha }$ components to get
restrictions on the arbitrary functions of $a_{\alpha \beta },b_{\alpha
\beta },$ and $c_{\alpha \beta}$.

From the $\binom{0}{0}$ component, we obtain,
\begin{equation}
\phi^2 (R^0_0 - \frac{1}{2}R) = 8\pi \phi \rho + \frac{1}{2}\omega(\partial_t \phi)^2 - \frac{1}{2}\phi K \partial_t \phi,
\label{BD00hom}
\end{equation}%
hence, the energy density satisfies,
\begin{equation}
8\pi \rho = [\phi_0 (\frac{1}{2}P_0 + \frac{1}{8}b^2 - \frac{1}{8} b^\beta_\alpha b^\alpha_\beta) - \frac{1}{2}\omega \frac{(\phi_1)^2}{\phi_0} + \frac{1}{2}\phi_1 b] + O(t^{n-1}).
\label{seriesdensityhom}
\end{equation}%
From the trace of the $\binom{\alpha}{\beta }$ component we have,
\begin{equation}
\phi^2 (R^\alpha_\alpha - \frac{3}{2}R) = 8\pi \phi T^\alpha_\alpha + \omega(\partial^\alpha \partial_\alpha \phi - \frac{3}{2}\partial^k \partial_k \phi) + \phi(\nabla^\alpha \nabla_\alpha \phi - 3\Box_g \phi),
\label{BDaahom}
\end{equation}
which gives the relation for the pressure,
\begin{equation}
8\pi p = [-\frac{1}{3}n(n-1)\phi_0 c - n(n-1)\phi_n]t^{n-2} + O(1).
\label{seriespressurehom}
\end{equation}%
The $\binom{\alpha }{\beta }$ components of the Brans-Dicke equations give,
\begin{equation}
\phi^2 (R^{\beta }_{\alpha} - \frac{1}{2}\delta^\beta_\alpha R) = -8\pi \phi p \delta^\beta_\alpha - \frac{1}{2}\omega(\partial_t \phi)^2 \delta^\beta_\alpha + \phi(\frac{1}{2}K^\beta_\alpha \partial_t \phi - \delta^\beta_\alpha {\partial_t}^2 \phi  - \frac{1}{2}\delta^\beta_\alpha K \partial_t \phi ),
\label{BDabhom}
\end{equation}%
and for the $(n-2)-$ order terms we find,
\begin{equation}
-\frac{1}{2}n(n-1)(\phi_0)^2(c^\beta_\alpha - c\delta^\beta_\alpha) = [\frac{1}{3}n(n-1)(\phi_0)^2 c + n(n-1)\phi_0 \phi_n]\delta^\beta_\alpha - n(n-1)\phi_0 \phi_n \delta^\beta_\alpha,
\end{equation}%
and therefore we have the constraints,
\begin{equation}
c^{\beta }_{\alpha }=\frac{c}{3}\delta ^{\beta }_{\alpha}.
\label{eqchom}
\end{equation}%
Equations (\ref{eqchom}) represent six relations between the initial data $c_{\alpha \beta}$ from which only one, out of the six, components of $c_{\alpha \beta}$ is arbitrary. Additionally, from the $\binom{0}{\alpha }$ components,
\begin{equation}
\phi R^0_\alpha = 8\pi (\rho + p)u_\alpha,
\label{BD0ahom}
\end{equation}%
we find three more relations for the velocities,
\begin{equation}
u_{\alpha }=-\frac{3\phi_0}{2n(n-1)(\phi_0 c + 3\phi_n)} (\nabla_\beta b^\beta_\alpha - \nabla_\alpha b)t^{2-n}.
\label{velocitieshom}
\end{equation}
Moreover, using the relation (\ref{Boxphi}), we find that,
\begin{equation}
{\partial_t}^2 \phi + \frac{1}{2}K\partial_t \phi = \frac{8\pi}{3 + 2\omega}(\rho - 3p).
\end{equation}
Taking into consideration relations (\ref{seriesdensity}) and (\ref{seriespressure}) the following equation,
\begin{equation}
\phi_n = \frac{\phi_0 c}{2\omega}.
\label{eqphi_nhom}
\end{equation}
Taking now into account the conservation laws $\nabla_i T^i_j=0$, the $(n-2)-$order term of the time component gives,
\begin{equation}
2\omega \phi_1 \phi_n - \phi_0 \phi_1 c = 0,
\end{equation}%
which is implied by (\ref{eqphi_nhom}).
For the same order, the spatial part of the stress-energy conservation is expressed as,
\begin{equation}
\frac{1}{2}n(n-1)\phi_0 (\nabla_\beta c^\beta_\alpha - \nabla_\alpha c) + \frac{1}{3}n(n-1)\phi_0 \nabla_\alpha c = 0,
\end{equation}%
namely that,
\begin{equation}
\frac{1}{6}\phi_0 (\nabla_\beta c^\beta_\alpha - \frac{1}{3} \nabla_\alpha c)= 0,
\end{equation}%
which is also implied by  Eq. (\ref{eqchom}).

Thus relations (\ref{seriesdensityhom}), (\ref{seriespressurehom}), (\ref{eqchom}), (\ref{velocitieshom}) and (\ref{eqphi_nhom}) give $6+6+1=13$ independent functions from the initial data $(a_{\alpha \beta}, b_{\alpha \beta}, c_{\alpha \beta})$, plus $1+1+0=2$ independent functions from $(\phi_0, \phi_1, \phi_n)$. Therefore, in total we have found $15$ independent functions. Subtracting the $4$ coordinate covariances we find $11$ independent arbitrary functions of the three space coordinates on a surface of constant $t$ time.

\end{document}